\documentclass[aps,prd,twocolumn,groupedaddress]{revtex4-2}

\usepackage{amsmath}
\usepackage{amssymb}
\usepackage{verbatim}
\usepackage{graphicx}
\usepackage{booktabs}
\usepackage{array}
\usepackage[dvipsnames]{xcolor}
\usepackage{xspace}
\usepackage{ulem}
\usepackage{lipsum}
\usepackage{xfrac}
\usepackage[colorlinks=true, allcolors=blue]{hyperref}
\usepackage{ulem}

\newcommand{\class}{\texttt{class}}
\newcommand{\mochiclass}{\texttt{mochi\_class}}
\newcommand{\hiclass}{\texttt{hi\_class}}

\newcommand{\lcdm}{$\Lambda$CDM}

\newcommand{\Dkin}{D_{\rm kin}}

\newcommand{\Mpl}{M_\ast^2}
\newcommand{\alphaM}{\alpha_{M}}
\newcommand{\alphaB}{\alpha_{B}}
\newcommand{\alphaK}{\alpha_{K}}
\newcommand{\cs}{c_{s}^2}
\newcommand{\csnum}{c_{s{\rm N}}^2}
\newcommand{\rhophi}{\rho_\phi}
\newcommand{\rhom}{\rho_{\rm m}}

\newcommand{\presm}{p_{\rm m}}
\newcommand{\presphi}{p_{\phi}}

\newcommand{\alphaBic}{\alpha_{B0}}

\newcommand{\Om}{\Omega_{\rm m}}
\newcommand{\diff}{{\rm d}}

\begin{document}


\title{Non-parametric exploration of minimally coupled gravity with phantom crossing}


\author{Matteo Cataneo}
\email[]{mcataneo@uni-bonn.de}
\affiliation{
$^{1}$Argelander-Institut für Astronomie, Universität Bonn, Auf dem Hügel 71, D-53121 Bonn, Germany \\
$^{2}$Ruhr University Bochum, Faculty of Physics and Astronomy, Astronomical Institute (AIRUB), German Centre for Cosmological Lensing, 44780 Bochum, Germany
}
\author{Kazuya Koyama}
\affiliation{
$^{3}$Institute of Cosmology and Gravitation, University of Portsmouth, Dennis Sciama Building, Burnaby Road, Portsmouth, PO1 3FX, United Kingdom \\
$^{4}$ Kavli IPMU (WPI), UTIAS, The University of Tokyo, Kashiwa, Chiba 277-8583, Japan\\
$^{5}$ Yukawa Institute for Theoretical Physics, Kyoto University, Kyoto 606-8502, Japan}

\date{\today}

\begin{abstract}
Recent measurements of the baryon acoustic oscillations by the Dark Energy Spectroscopic Instrument (DESI), especially when combined with cosmic microwave background (CMB) and supernova data, favor a late-time dark energy equation of state that crosses $w=-1$, which has been argued to point toward non-minimal conformal coupling in Horndeski gravity. We test this interpretation by performing a non-parametric exploration of the minimally coupled, luminal Horndeski subclass known as kinetic gravity braiding (KGB). Using \mochiclass{} and its manifestly stable effective field theory (EFT) basis implementation, we efficiently scan a broad class of models in which the EFT functions are allowed to vary freely in time, while enforcing the absence of ghost and gradient instabilities from the outset. We identify a set of KGB models that realize phantom crossing and remain broadly consistent with current probes of the background and linear large-scale structure, including CMB temperature and lensing power spectra, redshift-space distortions, cosmic shear, and the cross-correlation between galaxies and the integrated Sachs–Wolfe effect. Our results demonstrate that viable phantom-crossing solutions exist without conformal coupling, motivating future full Bayesian analyses of this model class with non-parametric EFT priors.
\end{abstract}


\maketitle

\section{Introduction}

The standard cosmological model, \lcdm, continues to provide an excellent phenomenological description of a wide range of observations~\cite{Scott2018,Ellis2018,Blanchard2025}, yet several recent datasets have begun to suggest possible departures from this framework~\cite{Abdalla2022,Peebles2022}. Most notably, the DR1 and DR2 baryon acoustic oscillation (BAO) analyses from the DESI collaboration~\cite{DESICollaboration2024,DESICollaboration2025}, when combined with external probes, indicate a $3\sigma$–$4\sigma$ preference for a dark energy equation of state that evolves across the boundary $w = -1$ at late times. Although this result is based on the Chevallier-Polarski-Linder (CPL) parameterization of the dark energy equation of state, more general parameterizations or reconstructions of the equation of state lead to a similar conclusion \cite{Lodha2024,Lodha2025, Gu2025}.
This picture is reinforced by the recent reanalysis of the Dark Energy Survey (DES) Y5 Type Ia supernova (SNIa) sample~\cite{Popovic2025}, which brings it into closer agreement with other SNIa compilations and likewise exhibits a persistent $\sim 3\sigma$ indication of phantom crossing. If confirmed, such behavior would point to new degrees of freedom in the dark sector and, crucially, to couplings between those degrees of freedom and gravity.

From a theoretical perspective, achieving phantom crossing in minimally coupled single–scalar-field models inevitably leads to ghost instabilities~\cite{DeFelice2011}. This has motivated a renewed interest in scalar–tensor theories in which the scalar field is conformally coupled to the metric. Recent works have shown that, when combined with CMB and SNIa data, the DESI BAO measurements appear to favor a non-zero conformal coupling at roughly the $2\sigma$ level~\cite{Ye2025,Pan2025,Wolf2025b}. In particular, Ref.~\cite{Ye2025} emphasizes that conformal coupling is the only single operator capable, by itself, of enabling phantom crossing.

An alternative viewpoint is that the late-time data may instead be pointing toward scalar–tensor theories with vanishing conformal coupling but non-trivial derivative interactions. In particular, phantom crossing can be stabilized by augmenting a k-essence Lagrangian with a cubic derivative self-interaction term~\cite{Deffayet2010}, giving rise to the class of models commonly referred to as kinetic gravity braiding (KGB). These theories generalize the original Cubic Galileon cosmology~\cite{Deffayet2009}, but typical Lagrangian formulations impose shift symmetry in the scalar field, which can strongly restrict the ability of the background equation of state to cross $w=-1$~\cite{Tsujikawa2025}. Moreover, such models often predict a time evolution of the metric potentials that leads to a negative cross-correlation between the galaxy density field and the integrated Sachs–Wolfe (ISW) effect~\cite{Kable2022}, in tension with observations. Breaking shift symmetry enlarges the space of viable background histories and can more easily allow for phantom crossing~\cite{Tsujikawa2025,Wolf2025c}, but whether these models can simultaneously satisfy constraints from the galaxy-ISW cross-correlation remains an open question.

The goal of this work is to investigate, within the framework of the effective field theory (EFT) of dark energy, whether KGB theories can provide viable alternatives to conformally coupled scenarios in light of current data. Rather than starting from a particular Lagrangian ansatz or a fixed low-dimensional parameterization of the EFT functions, we instead explore a broad, non-parametric family of models and impose theoretical stability from the outset. Our focus is on identifying minimally coupled cosmologies that remain consistent with a representative set of observables probing both the background expansion and the late-time growth of structure, including galaxy-ISW cross-correlations.

To carry out this exploration we use \mochiclass~\cite{Cataneo2024}, an extension of \hiclass~\cite{Zumalacarregui2016,Bellini2019} built on the Einstein-Boltzmann solver \class~\cite{Lesgourgues2011c,Blas2011}.
\mochiclass{} introduces two key features that greatly facilitate the search for viable models.
First, it implements a manifestly stable EFT basis~\cite{Lombriser2018}, ensuring by construction the absence of ghost and gradient instabilities at the linear level. Second, it allows these EFT functions and the background evolution to be specified as arbitrary arrays rather than simple analytic forms, thereby decoupling the exploration of theory space from any particular choice of Lagrangian or parametric template. Together, these capabilities enable an efficient, physically informed scan of the KGB landscape.

The remainder of this paper is organized as follows.
In Sec.~\ref{sec:background} we summarize the EFT description of luminal Horndeski gravity, introduce the KGB subclass, review both the stable-basis formulation used in \mochiclass{} and linear perturbation theory in modified gravity (MG).
Sec.~\ref{sec:methodology} describes our strategy for generating and selecting viable KGB models, and outlines the set of late-time observables and summary statistics considered in this work.
Our main results are presented in Sec.~\ref{sec:results}, where we examine the background evolution, the growth of perturbations, and predictions for CMB, redshift-space distortions (RSD), cosmic shear, and galaxy-ISW cross-correlations.
We conclude in Sec.~\ref{sec:conclusions} with a summary of our findings and an outlook on future developments.

Throughout this paper we work in units where $M_{\rm Pl}^2 \equiv 1/(8\pi G) = 1$ and $c = 1$, unless stated otherwise.
The standard cosmological parameters are fixed to the marginalized \lcdm{} posterior means from the 12.5HMcl TTTEEE likelihoods of Ref.~\cite{Efstathiou2021}, with the exception of the Hubble constant and total matter density, which we set to the mean values obtained in the joint DESI + CMB + DES~Y5 analysis of the $w_0w_a$CDM model in Ref.~\cite{DESICollaboration2025}:
$\omega_{\rm b} = 0.02226$, $\omega_{\rm c} = 0.11987$, $\tau_{\rm reio} = 0.0533$, $n_s = 0.9671$, $\ln(10^{10} A_s) = 3.04$, $h = 0.6674$.
We assume a flat Friedmann–Lemaître–Robertson–Walker (FLRW) background with metric signature $(-,+,+,+)$.

\section{Background}\label{sec:background}
\subsection{Horndeski gravity}
The most general local, Lorentz-invariant scalar-tensor theory in four dimensions that yields second-order equations of motion and predicts luminal gravitational wave propagation without fine-tuning is described by the following action~\cite{Horndeski1974,Deffayet2009a,Baker2017,Creminelli2017,Sakstein2017,Ezquiaga2017}:
\begin{subequations}
\begin{equation} \label{eq:full_horndeski}
  S=\int\mathrm{d}^{4}x\,\sqrt{-g}\left[\sum_{i=2}^{4}{\cal L}_{i} + \mathcal{L}_{\rm m}\right],
\end{equation}
\begin{align}
  &{\mathcal{L}}_{2} = G_{2}(\phi,\,X)\,, \\
  &{\mathcal{L}}_{3} = -G_{3}(\phi,\,X)\Box\phi\,, \\
  &{\mathcal{L}}_{4} = G_{4}(\phi)R\,. \label{eq:G4}
\end{align}
\end{subequations}
Here, $g$ is the determinant of the metric tensor $g_{\mu\nu}$, $\mathcal{L}_{\rm m}$ is the matter Lagrangian, and $G_{2,\,3,\,4}$ are arbitrary functions of the scalar field $\phi$ and its canonical kinetic term $X \equiv -\nabla^\mu \phi \nabla_\mu \phi/2$.

When restricted to the FLRW background and linear perturbations around it, EFT methods provide a complete description of the Horndeski Lagrangian in Eq.~\eqref{eq:full_horndeski} in terms of a small number of time-dependent functions~\cite{Frusciante2019}. \citet{Bellini2014} introduced an EFT formalism that explicitly separates the contributions of the scalar field to the background from those that affect the growth of structure, parameterized by the basis functions
\begin{align}\label{eq:alphas}
    & \{\rhophi,\,\alphaK,\,\alphaB,\,\,M_\ast^2\}\,,
\end{align}
where $\rhophi$ denotes the background energy density of the scalar field, $\alphaK$ (kineticity) captures its independent dynamics, $\alphaB$ (braiding) sources dark energy clustering, and $\Mpl$ is the effective Planck mass, which determines the cosmological strength of gravity and encodes the conformal coupling of the scalar field to the metric. In addition, the running, $\alphaM \equiv \frac{\diff \ln M_\ast^2}{\diff \ln a}$, encapsulates the time-dependence of the Planck mass. For models with $\rhophi \neq 0$, the background evolution can equivalently be specified using the equation of state of the scalar field, $w_\phi = \presphi / \rhophi$, where $\presphi$ is the pressure of the scalar field.  

To ensure the absence of ghost and gradient instabilities, the following conditions must be satisfied~\cite{DeFelice2011}:
\begin{equation}\label{eq:stability_conditions}
    \Dkin > 0, \quad M_\ast^2 > 0, \quad \cs > 0 \, ,
\end{equation}
where the de-mixed kinetic term is defined as
\begin{equation}\label{eq:Dkin}
    \Dkin \equiv \alphaK + \frac{3}{2}\alphaB^2 \, ,
\end{equation}
and the sound speed of the scalar field fluctuations is given by
\begin{align}\label{eq:cs2}
    c_{\mathrm{s}}^{2} = \frac{1}{\Dkin} & \left[(2-\alphaB) \left(-{\frac{\dot{H}}{H^{2}}}+{\frac{1}{2}}\alphaB+\alphaM\right) \right. \nonumber \\ 
    & \qquad\qquad\qquad \left. -{\frac{\left(\rho_{\mathrm{m}}+p_{\mathrm{m}}\right)}{H^{2}M_{\ast}^{2}}}+{\frac{\dot{\alpha}_{B}}{H}}\right] \, ,
\end{align}
where overdots denote derivatives with respect to cosmic time, $H$ is the Hubble parameter, and $\rhom$ and $\presm$ are the background energy density and pressure of matter, respectively. 

Alternatively to the description provided by Eq.~\eqref{eq:alphas}, one can start from the functions in Eq.~\eqref{eq:stability_conditions}, thereby enforcing stability by construction~\cite{Kennedy2018,Lombriser2018}. The braiding parameter \( \alphaB \) is then obtained by integrating Eq.~\eqref{eq:cs2} with an initial condition \( \alphaBic \), and the kineticity \( \alphaK \) follows from Eq.~\eqref{eq:Dkin}. The key advantage of this approach is that both theoretical exploration and data analyses can be restricted to stable models, i.e., those free from ghost and gradient instabilities. This facilitates the identification of viable Horndeski cosmologies, which can otherwise be difficult to access when sampling directly over the $\alpha$-functions, as the stable subspace may have a complex geometry or consist of multiple disconnected islands~\cite{Denissenya2018,Thummel2025}.

Although theoretically less problematic, mathematical (or classical) instabilities~\cite{Hu2014} can still affect the observational viability of Horndeski models. These instabilities manifest as exponentially growing modes in the perturbations and can be identified by examining the time- and scale-dependent coefficients in the equations of motion for the scalar field fluctuations. Although such models are typically ruled out by measurements of the late-time growth of structure, these additional stability criteria can be useful for tightening prior ranges in statistical analyses~\cite{Stlzner2025}, accelerating computations, and guiding theoretical exploration toward regions of parameter space consistent with observations. In this work, we monitor the growth rate of scalar field fluctuations during numerical integration and discard a model as soon as an instability with rate exceeding the Hubble constant, $H_0$, is detected.

Recent analyses have shown a preference for Horndeski models with non-minimal conformal coupling—that is, with $\alphaM \neq 0$—when fitting DESI BAO data~\cite{Ye2025, Chudaykin2024, Wolf2025a, Wolf2025b, Pan2025}. However, as highlighted by \citet{Wolf2025c} and \citet{Tsujikawa2025}, cubic Galileon models with broken shift symmetry can produce a dark energy equation of state that crosses the phantom divide while remaining consistent with data probing the late-time growth of structure (see Ref.~\cite{Linder2025} for a detailed discussion on shift-symmetric Horndeski cosmologies). These models belong to the broader class of KGB theories~\citep{Deffayet2010}, a subset of Horndeski gravity where the scalar field is minimally coupled to gravity—i.e., in Eq.~\eqref{eq:G4}, $G_4(\phi) = 1/2$, which corresponds to setting $M_\ast =1$.

Here, we adopt a complementary phenomenological approach to that of Ref.~\cite{Wolf2025c}. Instead of specifying a particular Lagrangian, we explore the space of viable KGB models by directly sampling the stable basis functions $\{ \Dkin, \cs \}$, while assuming a CPL parameterization for the dark energy equation of state to parameterize $\rhophi$. We consider three representative cases: (i) the DESI DR2 + CMB + DES Y5 best-fit values $\{w_0, w_a\} = \{-0.75, -0.86\}$ in Ref.~\citep{DESICollaboration2025}; and (ii, iii) two mirage dark energy scenarios~\citep{Linder2007}, defined by the relation $w_a = -3.66(1 + w_0)$, with either $w_0 = -0.5$ or $w_0 = -0.85$. The latter values lie within the $2\sigma$ credible region from the combination DESI DR2 + CMB + Union3 in Ref.~\citep{Lodha2025}. Note that the recent reanalysis of DES Y5 SNIa shifted the best-fit values to  $\{w_0, w_a\} = \{-0.802, -0.72\}$ \cite{Popovic2025}. These values satisfy the relation imposed by the mirage dark energy and are within the two values we selected.

\subsection{Linear growth of structure}

The line element for the perturbed FLRW metric in the conformal Newtonian gauge, expressed in terms of conformal time $\tau$ and comoving coordinates $\mathbf{x}$, is given by
\begin{equation}
\diff s^2 = a^2(\tau)\left[-(1 + 2\Psi)\diff\tau^2 + (1 - 2\Phi)\diff\mathbf{x}^2\right] \, ,
\end{equation}
where $a$ is the scale factor, and the two metric potentials $\Psi$ and $\Phi$ describe, respectively, the Newtonian potential and the intrinsic spatial curvature perturbation.
While \mochiclass\ solves the full linearized Einstein-Boltzmann system for the metric, scalar field, and stress–energy perturbations, under the sub-horizon quasi-static approximation the effects of the scalar field on the growth of structure and gravitational lensing can be encapsulated by three functions of time and scale: the effective gravitational coupling for non-relativistic matter, $\mu$; the gravitational slip, $\gamma$; and the effective coupling for relativistic species, $\Sigma = \mu(1 + \gamma)/2$.
These functions modify the dynamics of matter perturbations through the generalized Poisson equations~\citep{Zhao2008}, which in Fourier space in the absence of anisotropic stress read
\begin{align}
k^2 \Psi &= -\frac{1}{2} \mu(a,k) a^2 \rhom \Delta_{\rm m} \, , \label{eq:newton_potential} \\
k^2 (\Phi + \Psi) &= -\Sigma(a,k) a^2 \rhom \Delta_{\rm m} \, , \label{eq:weyl_potential} \\
\Phi &= \gamma(a,k) \Psi \, , \label{eq:slip}
\end{align}
where $\Delta_{\rm m}$ denotes the comoving matter-density contrast.

In Horndeski gravity, the quasi-static functions take compact forms that depend on the background evolution and the EFT functions (see, e.g., Ref.~\cite{Pogosian2016}):
\begin{subequations}\label{eq:efe_qsa}
    \begin{align}
        \mu(k,a) &= \frac{1}{M_\ast^2} \frac{\mu_{\rm p} + k^2 \csnum M_\ast^2 \mu_{\infty}/a^2 H^2}{\mu_{\rm p} + k^2 \csnum/a^2 H^2} \, , \\
        \gamma(k,a) &= \frac{\mu_{\rm p} + k^2 \csnum M_\ast^2 \mu_{Z,\infty}/a^2 H^2}{\mu_{\rm p} + k^2 \csnum M_\ast^2 \mu_{\infty}/a^2 H^2} \, ,
    \end{align}
\end{subequations}
where $\csnum \equiv \Dkin \cs$, and expressions for $\mu_{\rm p}$, $\mu_\infty$ and $\mu_{Z,\infty}$ can be found in Appendix D of Ref.~\cite{Cataneo2024}. For KGB models, where $M_\ast = 1$, it follows that $\gamma = 1$, or equivalently, $\mu = \Sigma$.
Moreover, in models where the effective mass of the scalar field is comparable to the Hubble scale, the following sub-horizon approximation holds:
\begin{equation}\label{eq:mu_infinity}
    \mu(a) \approx \mu_\infty(a) = 1 + \frac{\alpha^2_{\rm B}}{2\csnum} \, .
\end{equation}
The positivity of this modification implies a faster growth of structure due to the presence of an attractive fifth force. Consequently, in such models, structure formation can proceed more slowly than in \lcdm{} only through changes to the background expansion. 

To study the large-scale structure phenomenology on linear scales, it is useful to compute the growth function $D$ by integrating
\begin{equation}\label{eq:growth}
D^{\prime\prime} + \frac{3}{2a}\left[1 - w_\phi(a) \Omega_\phi(a) \right] D^{\prime}
- \frac{3}{2} \frac{\Omega_{\rm m}(a)}{a^2} \mu(a) D = 0 \, ,
\end{equation}
where primes denote derivatives with respect to $a$, $\Omega_\phi$ is the energy density of the scalar field in units of the critical density. The initial conditions are set deep in the matter-dominated era: $D(a_{\rm ini}) = D^\prime(a_{\rm ini}) = a_{\rm ini}$. From this, the linear growth rate is given by
\begin{equation}\label{eq:growth_rate}
    f(a) = \frac{\diff \ln D}{\diff \ln a} \, ,
\end{equation}
which, when combined with the amplitude of matter clustering, provides a convenient summary statistic for the growth of structure: $f\sigma_8$~\cite{Song2009}. 
We use the public code \texttt{MGrowth}~\footnote{\href{https://github.com/MariaTsedrik/MGrowth}{https://github.com/MariaTsedrik/MGrowth}} to compute both the growth function and the growth rate for a given expansion history and choice of the modified gravity parameter $\mu(a)$. To provide a qualitative assessment of the viability of our models, we compare them, where appropriate, to the phenomenological scale-independent parameterizations~\cite{Simpson2015}:
\begin{eqnarray}\label{eq:mu_sigma_pheno}
\mu(a) &=& 1 + \mu_0 \frac{\Omega_{\phi}(a)}{\Omega_{\phi}} \, , \\
\Sigma(a) &=& 1 + \Sigma_0 \frac{\Omega_{\phi}(a)}{\Omega_{\phi}} \, ,
\end{eqnarray}
where $\mu_0$ and $\Sigma_0$ are free parameters that control the present-day departure from GR.

\section{Methodology}\label{sec:methodology}

\subsection{Generating minimally coupled models}\label{sec:generating_models}

Following the approach of \citet{Cataneo2024}, we parameterize the stable basis functions $\Dkin$ and $\cs$ as smooth deformations of power laws in the time variable $x \equiv \ln a$. Specifically, we write
\begin{align}
\Dkin &= \exp\left[ \zeta_D x + b_D + \mathcal{E}_D(x) \right], \label{eq:Dkin_params} \\
\cs &= C_{c_s} \exp\left[ \mathcal{E}_{c_s}(x) \right], \label{eq:cs2_params}
\end{align}
where $\zeta_D$, $b_D$, and $C_{c_{s}}$ are free parameters controlling the asymptotic early-time behavior, and the residual functions $\mathcal{E}_D$, $\mathcal{E}_{c_{s}}$ encode smooth, time-dependent deviations. Each $\mathcal{E}_Y$ is drawn from a Gaussian Process (GP) with zero mean and a squared exponential kernel,
\begin{equation}\label{eq:gp_kernel}
K(x, x’) = \sigma^2 \exp\left[ -\frac{(x - x’)^2}{2\lambda^2} \right],
\end{equation}
with variance $\sigma^2$ and characteristic length-scale $\lambda$ as hyperparameters. The GPs are conditioned such that $\mathcal{E}_Y \to 0$ as $x \ll 0$, ensuring the basis functions approach pure power laws deep in the matter-dominated era. The background expansion is specified independently via the dark energy equation of state $\{ w_0,w_a \}$, which sets the evolution of the scalar field energy density through the continuity equation. The initial condition for the braiding parameter \( \alphaBic \) is chosen such that the numerical solution of Eq.~\eqref{eq:cs2} converges to the early-time approximation~\cite{Cataneo2024}
\begin{align}\label{eq:braid_early}
    \alphaB \approx \, &\frac{C_{c_{s}}}{\zeta_D - 1/2}\exp\left[\zeta_D x + b_D\right] \nonumber \\ 
                &+ \left( \frac{1 - \Om}{\Om} \right) \frac{\zeta_\rho}{\zeta_\rho + 5/2} \exp\left[(\zeta_\rho + 3)x + b_\rho \right] \, ,
\end{align}
with
\begin{align}
    \zeta_\rho &= -3(1+w_0+w_a) \, , \\
    b_\rho &= -3w_a \, . 
\end{align}

Finally, we generate over 250\,000 KGB models by combining deviation functions drawn from the trained GPs—with variances $\sigma_D^2 = 50$, $\sigma_{c_{s}}^2 = 10$, and correlation lengths $\lambda \in \{0.9, 1, 1.1\}$—with power laws specified by parameters sampled from wide uniform distributions:
\begin{align}
\zeta_D &\sim \mathcal{U}(2, 15) , \\
b_D &\sim \mathcal{U}(-10, 5) , \\
C_{c_{s}} &\sim \mathcal{U}(0.1, 1.5) .
\end{align}
From the samples passing the mathematical stability conditions implemented in \mochiclass, we select four models (M1–M4) consistent with the DESI DR2 + CMB + DES Y5 best-fit background $\{w_0, w_a\} = \{-0.75, -0.86\}$, and two additional models with mirage dark energy backgrounds: $\{w_0, w_a\} = \{-0.5, -1.83\}$ (M5) and $\{-0.85, -0.549\}$ (M6). Selected models are required to satisfy two phenomenological criteria: (i) the matter power spectrum at redshift $z = 0$ must deviate by less than 3\% from \lcdm{} on sub-horizon scales, where this threshold is adopted as a phenomenological criterion to ensure broad consistency with current large-scale structure constraints; and (ii) the late-time ISW effect must be positive (see Sec.~\ref{sec:late_observables}). The six models selected for further analysis are intended as representative examples of the viable KGB parameter space, chosen to illustrate the range of phenomenological behaviors that arise under these criteria, rather than an exhaustive subset of all allowed solutions.

\subsection{Theoretical predictions for late–time observables}
\label{sec:late_observables}

Beyond assessing the impact of our KGB models on the primary CMB temperature anisotropies and the growth-rate combination $f\sigma_{8}$, we consider a set of projected late-time observables derived from cross-correlations of large-scale structure tracers. In particular, we compute angular power spectra involving the CMB lensing convergence, cosmic shear, the ISW contribution to the CMB temperature anisotropies, and the galaxy density field. These probes track the evolution of the gravitational potentials and thus provide crucial information on low-redshift modifications to the laws of gravity.

\paragraph*{General formalism.}  The angular cross–power spectrum of any two projected tracers, $A$ and $B
$, is computed as (see, e.g., Ref.~\cite{Chisari2019})
\begin{equation}
C_\ell^{AB} \;=\; 4\pi \int_0^\infty \frac{\mathrm{d}k}{k} \,\mathcal{P}_{\mathcal{R}}(k)\,
\Delta_\ell^A(k)\,\Delta_\ell^B(k)\,,
\end{equation}
where $\mathcal{P}_{\mathcal{R}}(k)$ is the dimensionless power spectrum of the primordial curvature perturbations, and $\Delta_\ell^A$ and $\Delta_\ell^B$ are the transfer functions for the corresponding tracers. Throughout this work we restrict to linear theory and compute matter and potential transfer functions using \mochiclass.

\paragraph*{CMB lensing convergence.}  The gravitational potential of the large-scale structure coherently deflects the CMB photons; this effect is captured by the lensing convergence $\kappa_{\rm CMB}$, which probes the Weyl (or lensing) potential $(\Phi+\Psi)/2$. Its transfer function is
\begin{align}
    \Delta_\ell^{\kappa_{\rm CMB}}(k) = -\frac{\ell(\ell + 1)}{2} \int_0^{z_\ast} &\frac{\diff z}{H(z)} \, \left( \frac{\chi_\ast - \chi}{\chi \chi_\ast} \right) \nonumber \\[1em]
    &\times T_{\Phi + \Psi}(k, z) j_\ell\!\left(k \chi\right) \,,
\end{align}
where $\chi(z)$ is the comoving distance, $\chi_\ast$ is the comoving distance to the last-scattering surface, and $T_{\Phi+\Psi}$ is the transfer function of the sum of the metric potentials.

\paragraph*{Galaxy number counts.}  The projected fluctuation of the galaxy number density is a biased tracer of the underlying matter density field. For cross-correlations with the CMB temperature field at redshifts $z \lesssim 1$, contributions from redshift-space distortions and magnification bias are negligible~\cite{Krolewski2025}, thus the transfer function reduces to
\begin{equation}
\Delta_\ell^{\rm g}(k)
= \int_0^\infty \mathrm{d}z\,n_{\rm g}(z)\,b(z)\,T_\delta(k,z)\,
j_\ell\!\left(k\,\chi\right)\,,
\end{equation}
where $n_{\rm g}(z)$ is the normalised galaxy redshift distribution, $b(z)$ is the linear galaxy bias, and $T_\delta$ is the linear matter transfer function. We adopt the redshift distribution of the unWISE blue sample from Ref.~\cite{Krolewski2025}. Although modified gravity alters the growth of structure, its impact on the galaxy bias is subdominant compared to the observational uncertainties in the galaxy-ISW cross-correlation (see Ref.~\cite{Renk2017}). We therefore use the effective bias inferred for this galaxy sample under the assumption of a \lcdm{} cosmology~\cite{Krolewski2025}.

\paragraph*{ISW temperature fluctuations.}  Time variation of the lensing potential at late times generates secondary CMB anisotropies known as the integrated Sachs–Wolfe effect. Since this is the only contribution to the CMB temperature field that correlates with galaxy number counts on large angular scales, the corresponding transfer function simplifies to
\begin{equation}
\Delta_\ell^T(k)
= \int_{0}^{z_\ast} \mathrm{d}z\,H(z)\,
\frac{\mathrm{d}}{\mathrm{d}z}\!\left[T_{\Phi+\Psi}(k,z)\right]
j_\ell\!\left(k\,\chi\right)\,.
\end{equation}
The sign of the galaxy-ISW cross-spectrum is determined by the time evolution of the lensing potential over the redshift range of the galaxy sample: it is positive if the potential decays, as in $\Lambda$CDM, and negative if it deepens (see, e.g., Ref.~\cite{Renk2017}).

\paragraph*{Cosmic shear.} Weak lensing by the large-scale structure also shears the shapes of distant galaxies. The corresponding transfer function is
\begin{align}\label{eq:kernel_shear}
    \Delta_{\ell,i}^{\gamma}(k) = -\frac{1}{2} \sqrt{\frac{(\ell+2)!}{(\ell-2)!}} 
                        &\int_{0}^{\infty} \frac{\diff z}{H(z)}\, W_{i}(z) \nonumber \\[1em] 
                        & \times T_{\Phi+\Psi}(k,z)\,j_\ell\!\left(k\,\chi(z)\right)\,,
\end{align}
with the lensing weight
\begin{equation}
W_{i}(z) = \int_{z}^{\infty} \mathrm{d}z'\,
n_{{\rm s},i}(z')\,
\left(\frac{\chi'-\chi}{\chi'\,\chi}\right)\,.
\end{equation}
Here $n_{{\rm s},i}(z)$ denotes the normalised redshift distribution of source galaxies in the $i$th tomographic bin. In this work we adopt the source distributions of the four DES~Y3 tomographic bins~\cite{Myles2021}.

\paragraph*{Observables considered.}  In summary, we compute the following angular spectra:
\begin{itemize}
\item The CMB lensing power spectrum, $C_\ell^{\phi\phi} = 4 C_\ell^{\kappa_{\rm CMB}\kappa_{\rm CMB}}/[\ell(\ell+1)]^2$, calculated directly with \mochiclass.
\item The cosmic shear power spectrum, $C_\ell^{\gamma\gamma}$, for the four DES~Y3 tomographic bins, computed with the Core Cosmology Library (\texttt{CCL})~\cite{Chisari2019} in the Limber approximation.
\item The galaxy-ISW cross-spectrum, $C_\ell^{T {\rm g}}$, also computed with \texttt{CCL} in the Limber approximation.
\end{itemize}

These three projected summary statistics are primarily sensitive to the lensing potential and complement the information provided by redshift-space distortions, summarised by $f\sigma_8$, which probes the Newtonian potential~\cite{Percival2008}.

\section{Results}\label{sec:results}

In this section we present a first, qualitative examination of the phenomenology of the selected KGB models. Our goal is not to perform parameter inference or optimize against current data, but rather to establish whether these models are broadly consistent with key late-time cosmological observables--thereby motivating a more rigorous statistical analysis in future work. The comparisons that follow should therefore be interpreted as an initial viability check, focused on the physical signatures that distinguish these models from standard dark energy scenarios.

\begin{figure*}
  \centering
  \begin{minipage}[t]{0.48\textwidth}
    \centering
    \includegraphics[height=0.33\textheight]{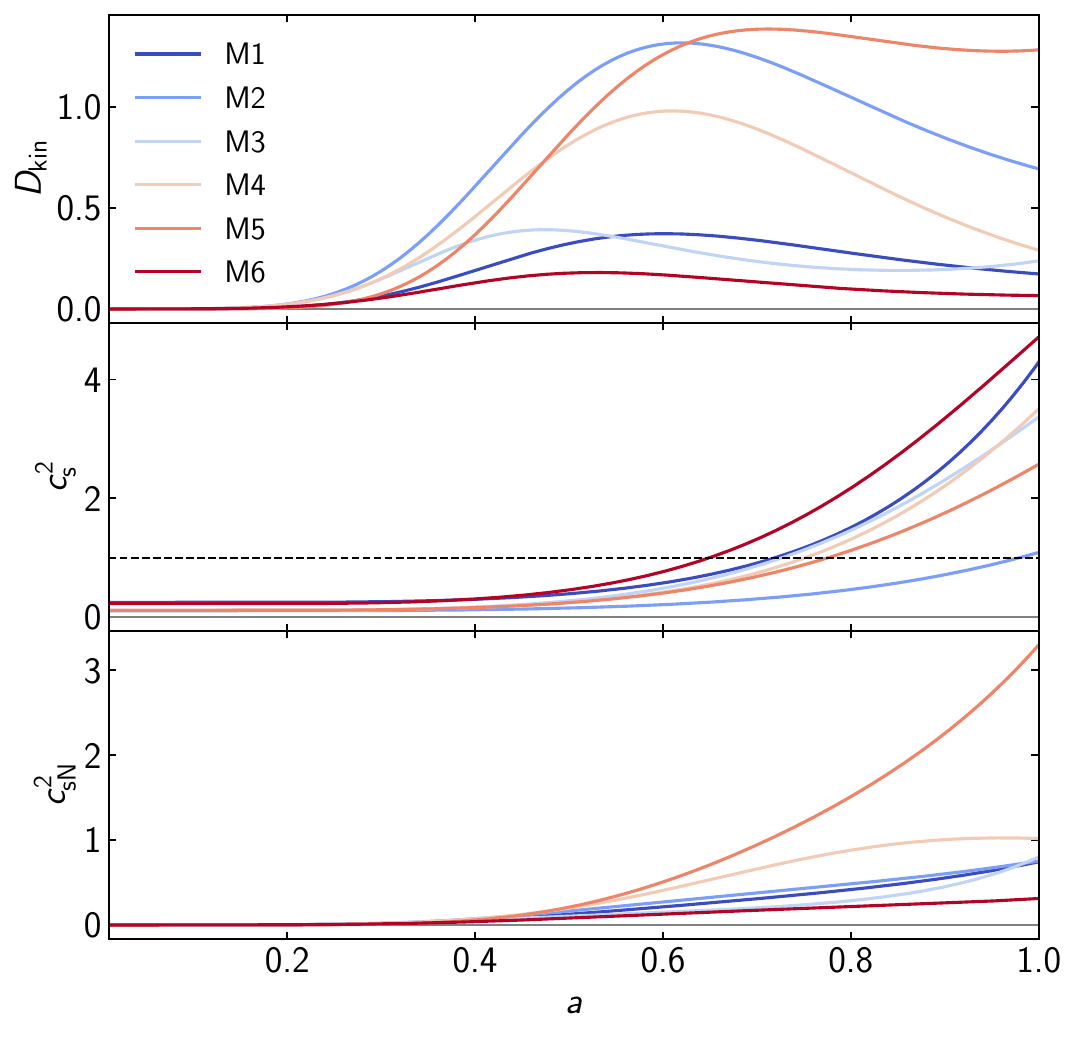}
  \end{minipage}%
  \hfill
  \begin{minipage}[t]{0.48\textwidth}
    \centering
    \includegraphics[height=0.33\textheight]{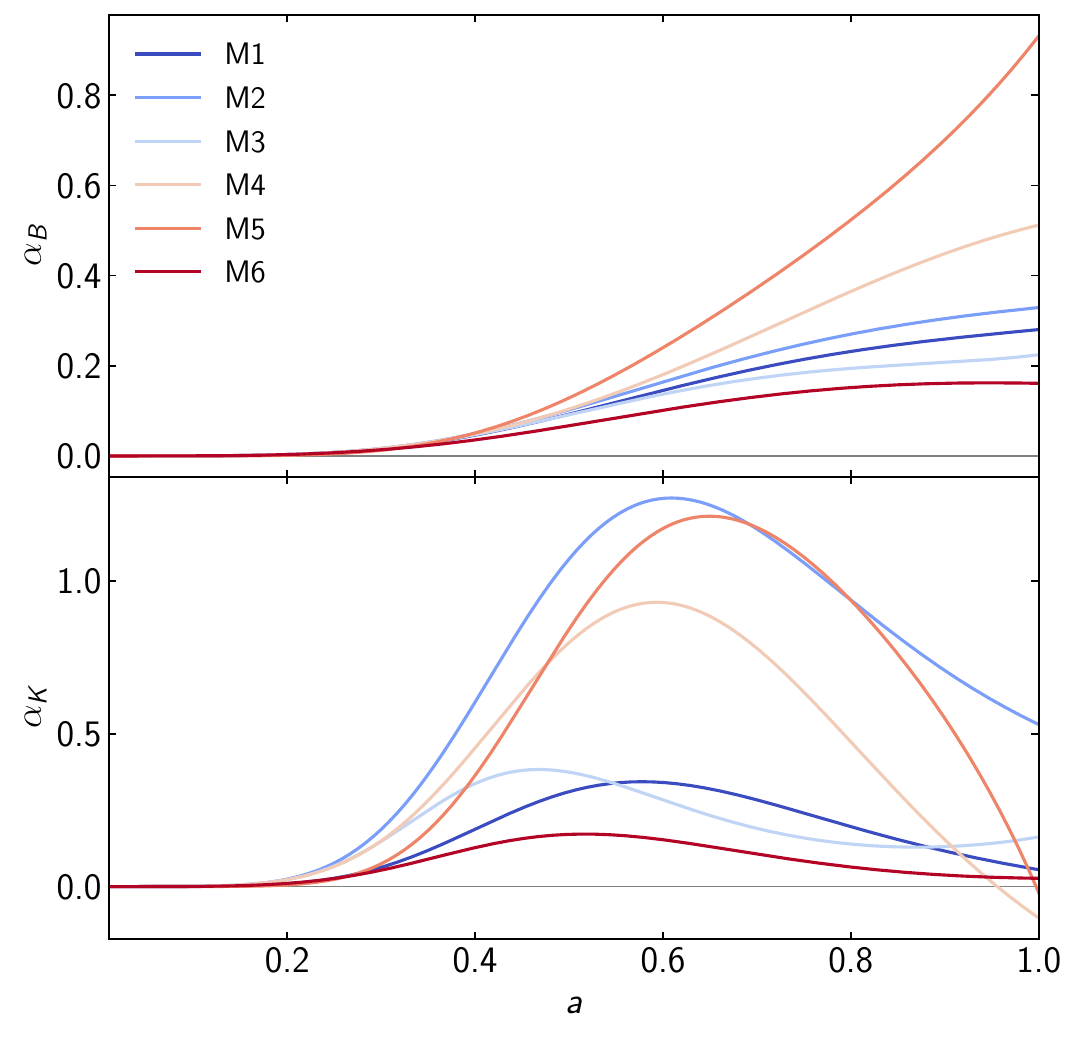}
  \end{minipage}
  \caption{\label{fig:basis_functions} {\it Left:} Evolution of the stable basis functions $\{\Dkin, \cs\}$ for the six selected KGB models. In all cases, the sound speed starts out subluminal ($\cs < 1$) and becomes superluminal ($\cs > 1$) at late times. For convenience, we also show their product, $\csnum$, which is more directly related to the modification of the Newtonian potential (see Eq.~\ref{eq:mu_infinity}). {\it Right:} Derived parameters braiding, $\alphaB$, and kineticity, $\alphaK$, for the same models shown in the left panel.} 
\end{figure*}
\begin{figure*}
  \centering
  \begin{minipage}[t]{0.48\textwidth}
    \centering
    \includegraphics[height=0.33\textheight]{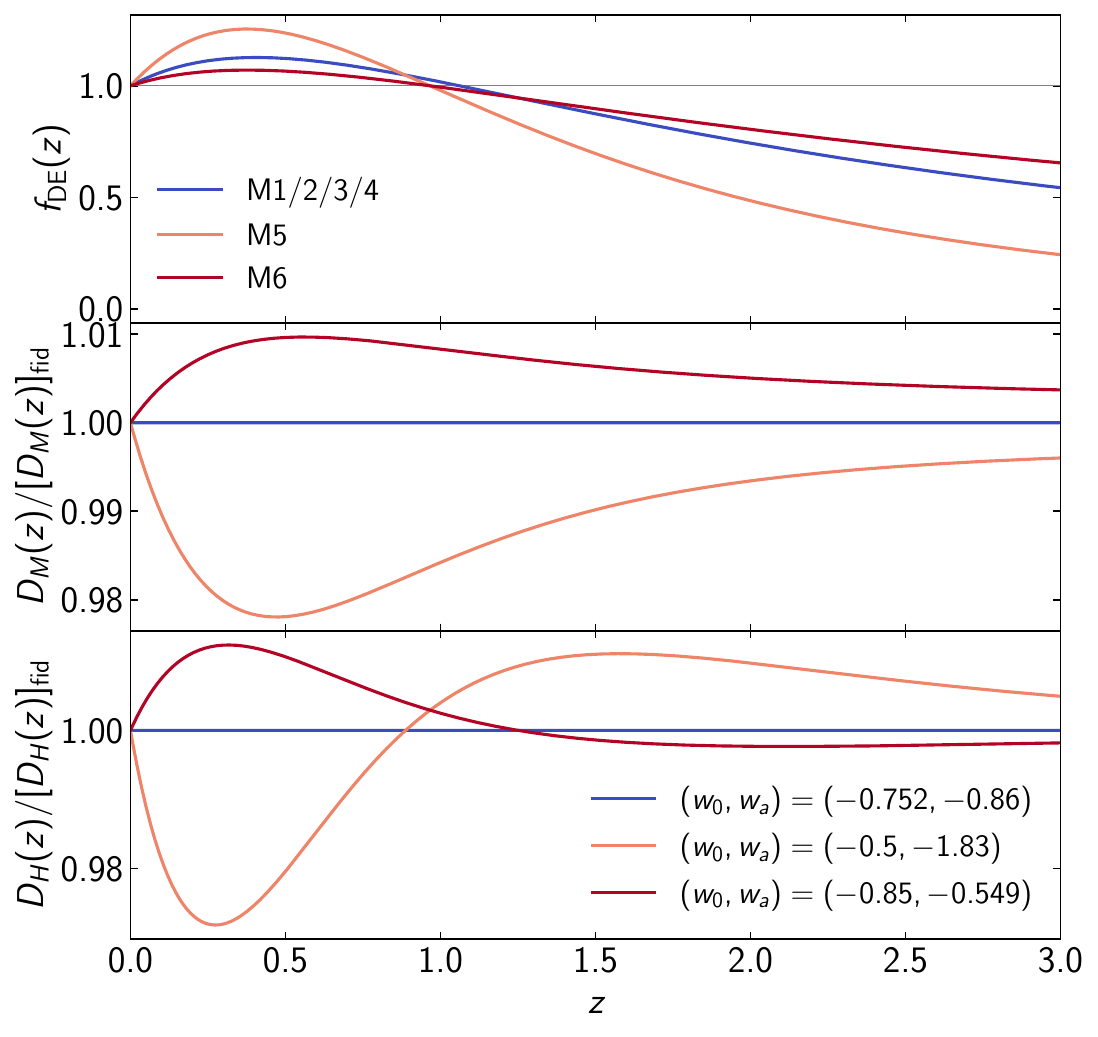}
  \end{minipage}%
  \hfill
  \begin{minipage}[t]{0.48\textwidth}
    \centering
    \includegraphics[height=0.33\textheight]{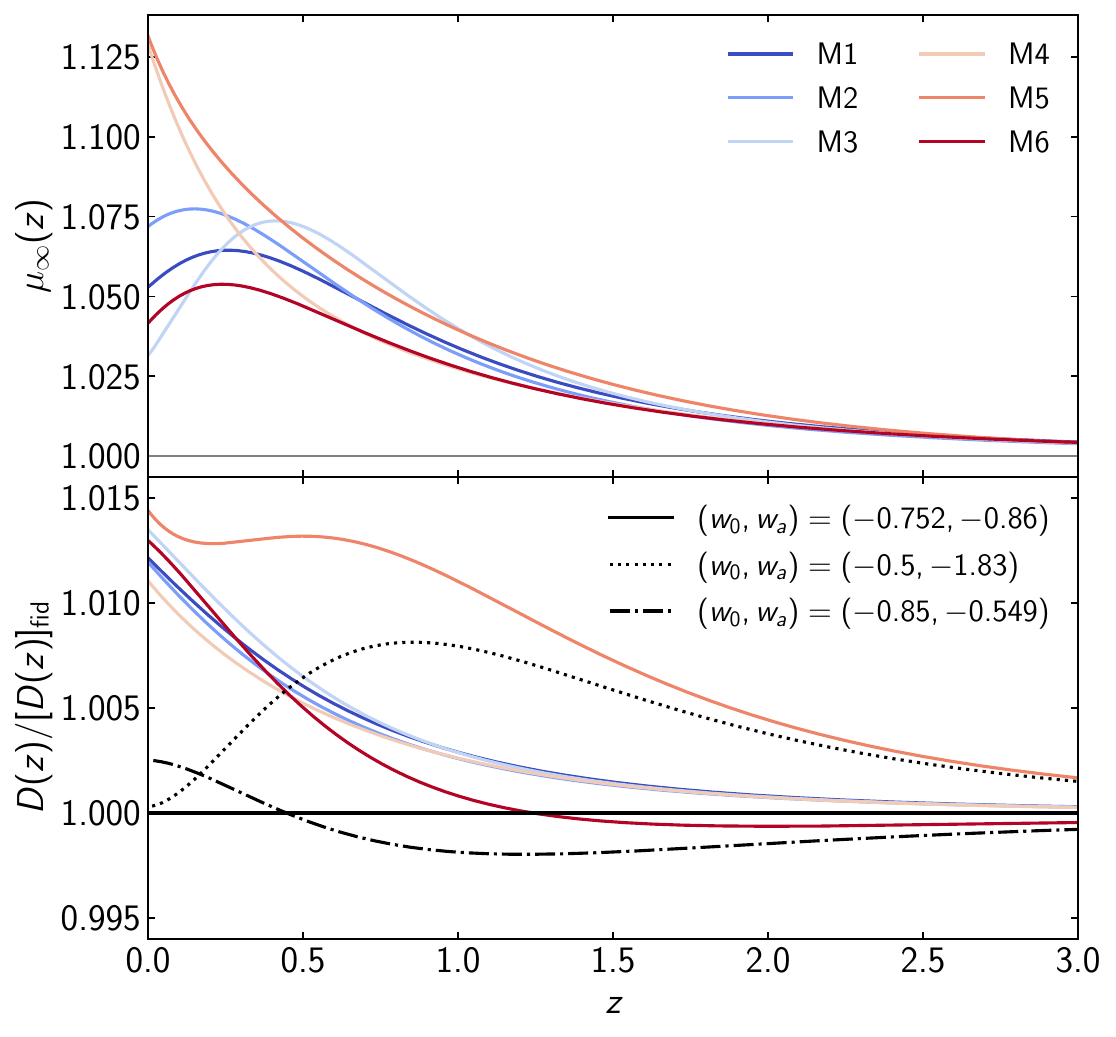}
  \end{minipage}
  \caption{\label{fig:background_perturbations} {\it Left:} Evolution of background quantities for the six selected KGB models. From top to bottom we show the present-day normalized dark energy density $f_{\rm DE}$, the transverse comoving distance $D_{M}$ relative to the DESI DR2 best-fit $w_{0}w_{a}$CDM prediction, and the corresponding Hubble distance $D_{H}$ relative to the same fiducial model. Models M1–M4 share this best-fit expansion history (blue), while M5 and M6 follow the two mirage dark energy backgrounds (orange and red). {\it Right:} Linear-theory predictions for the quasi-static modification of the Newtonian potential, $\mu_{\infty}$ (top), and for the growth function, $D(z)$, relative to the DESI DR2 best-fit $w_{0}w_{a}$CDM model (bottom). Coloured curves correspond to the six KGB models M1–M6, while black lines show the three reference $w_{0}w_{a}$CDM cosmologies used throughout this work.}
\end{figure*}

Figure~\ref{fig:basis_functions} illustrates the evolution of the stable basis functions $\{ \Dkin,\cs \}$ for the six selected KGB models, together with the derived parameters $\alphaB$ and $\alphaK$. Although the individual components $\Dkin$ and $\cs$ differ substantially across models, their product $\csnum$ exhibits a remarkably similar evolution for M1–M3. This in turn leads to similar braiding functions through Eq.~\eqref{eq:cs2}, since these models share an identical expansion history (see Fig.~\ref{fig:background_perturbations}, left panel). While all models become superluminal at late times ($\cs > 1$), this feature does not necessarily imply causality violation in scalar–tensor theories (see, e.g., Refs.~\cite{Bruneton2006,Babichev2007}).

By contrast, the kineticity parameter $\alphaK$ displays a wide range of non-trivial time evolutions. Although this parameter does not directly affect the growth of structure on sub-horizon scales, it plays a crucial role in ensuring physical stability within the $\alpha$-parameterization. Importantly, the behaviors observed here cannot be captured by commonly used ansätze, such as $\alphaK \propto \Omega_{\rm DE}$. Starting instead from generic $\{ \alphaB,\alphaK \}$ parameterizations would likely result in a large fraction of models suffering from gradient instabilities ($\cs < 0$), making the physically viable configurations explored here exceedingly difficult to discover without the stable basis construction.

The left panel of Fig.~\ref{fig:background_perturbations} shows the background evolution for the selected KGB models. The top panel displays the energy density of the scalar field normalized to its present-day value, $f_{\rm DE}(z) \equiv \rho_{\phi}(z)/\rho_{\phi0}$, illustrating that M1–M4 share an identical expansion history by construction, while M6 remains closest to the $\Lambda$CDM behavior and M5 shows the largest departure. The middle and bottom panels show the transverse comoving distance and the Hubble distance~\footnote{Here we explicitly reintroduce the speed of light for clarity.},
\begin{equation}
D_{M}(z) = c \int_0^z \frac{\mathrm{d}z'}{H(z')}\,,
\qquad
D_{H}(z) = \frac{c}{H(z)}\,,    
\end{equation}
respectively, each normalized to the DESI DR2 best-fit $w_0w_a$CDM prediction. Over the full redshift range relevant to current observations, the deviations from the fiducial model in these distance measures remain below 3\%, highlighting that the expansion history alone provides only limited discriminating power among these scenarios.

The right panel of Fig.~\ref{fig:background_perturbations} shows the impact of modified gravity on structure formation through the quasi–static modification of the Newtonian potential, $\mu_{\infty}(z)$, and the linear growth function $D(z)$ (normalized to the DESI DR2 best-fit $w_0w_a$CDM prediction). Note that for the calculation of the growth function and observables in the $w_0w_a$CDM cosmologies, we assume that perturbations are not modified, $\mu = \Sigma=1$. The different KGB models exhibit two distinct behaviors in the effective gravitational coupling: M4 and M5 feature a monotonically increasing fifth force, while M1–M3 and M6 instead show a maximum around $z \approx 0.25$–0.6. Despite variations as large as 10\% at low redshift for M1–M4, these differences translate into a sub-percent variation in the growth function over the same interval. This reflects the fact that growth is sensitive to the cumulative effect of the modified gravitational strength over cosmic time. At $z \gtrsim 3$, all models approach the General Relativity (GR) limit and converge to their respective reference $w_0w_a$CDM cosmologies. For all models the growth of structure is enhanced relative to the fiducial cosmology, with the only exception of M6 at $z \gtrsim 1.5$ owing to its background evolution.

\begin{figure}
  \centering
  \includegraphics[width=\columnwidth]{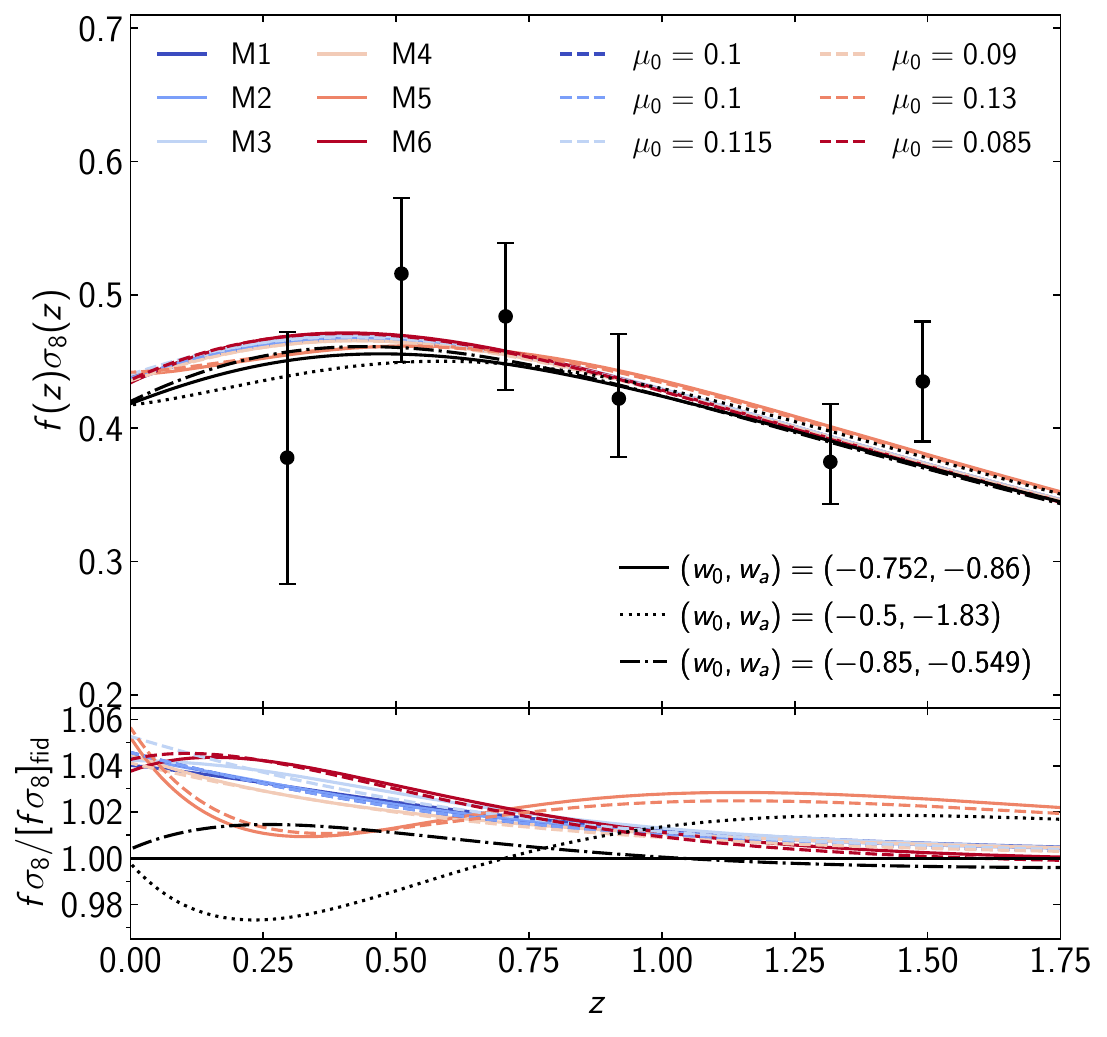}
  \caption{\label{fig:fsigma8} {\it Top:} Predictions for the RSD parameter combination $f\sigma_{8}$ for the six KGB models M1–M6 (solid colored lines), along with their matched phenomenological  models parameterized by $\mu_{0}$ (see Eq.~\ref{eq:mu_sigma_pheno}; dashed lines) and the three $w_{0}w_{a}$CDM cosmologies (black lines). Black points with error bars show the DESI DR1 measurements reported in Fig. 14 of Ref.~\cite{Adame2025}. {\it Bottom:} Ratio of the predicted $f\sigma_8$ for each model to the DESI DR2 best-fit $w_0w_a$CDM cosmology. Line styles and colors are the same as in the top panel. 
  }
\end{figure}

The predictions for the parameter combination $f\sigma_{8}$ are shown in Fig.~\ref{fig:fsigma8} alongside the DESI DR1 compressed RSD measurements (black points) from Figure~14 of Ref.~\cite{Adame2025}. As noted in that work, some care is required when interpreting these data, as they should ideally be adjusted for the effects of the fiducial cosmology assumed during the compression process. With this caveat in mind, all of our selected KGB models lie well within the observational uncertainties across the full redshift range, suggesting that current RSD data impose only mild constraints on this class of theories. Moreover, the matched phenomenological models parameterized by $\mu_0$ closely track the full KGB predictions for $z < 1.5$, confirming that a gravitational coupling proportional to the background scalar-field energy density accurately captures the relevant fifth-force effects. The fitted values of $\mu_0$ are in good agreement with the constraint $\mu_0 = -0.24^{+0.32}_{-0.28}$ from Ref.~\cite{Ishak2025}.

\begin{figure*}
  \centering
  \begin{minipage}[t]{0.55\textwidth}
    \centering
    \includegraphics[height=0.3\textheight]{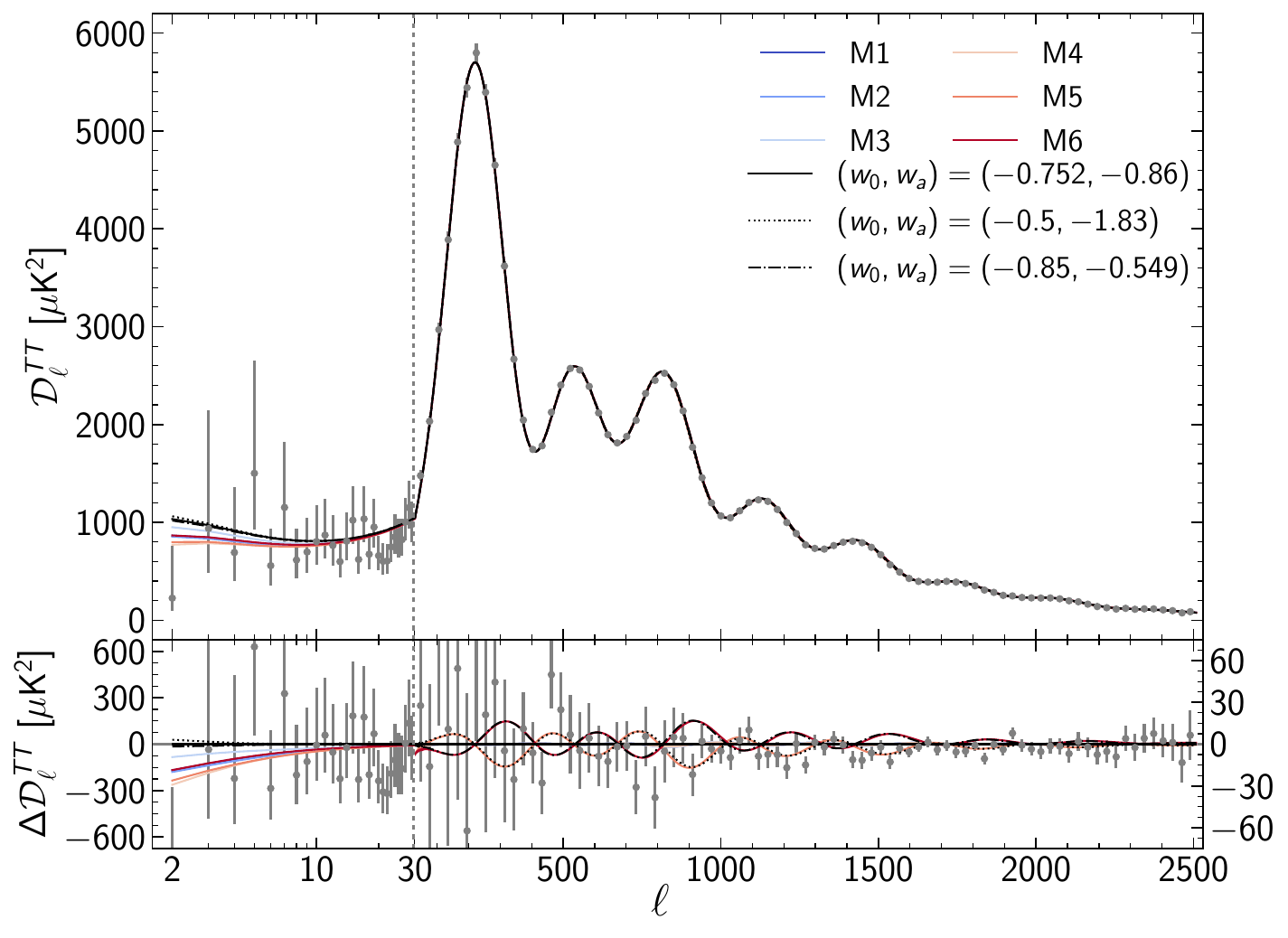}
  \end{minipage}%
  \hfill
  \begin{minipage}[t]{0.45\textwidth}
    \centering
    \includegraphics[height=0.3\textheight]{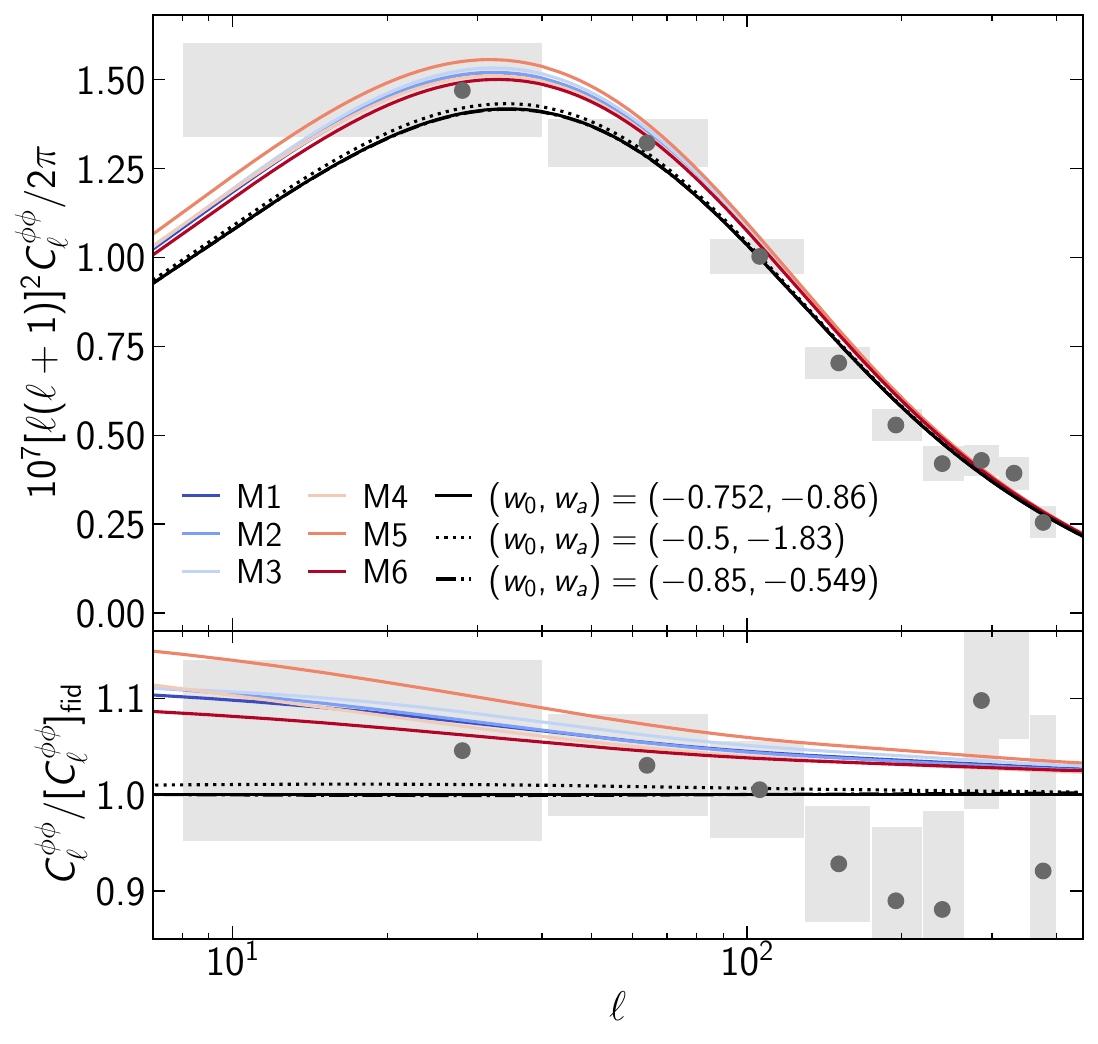}
  \end{minipage}
  \caption{\label{fig:cmb} \textit{Left:} CMB temperature power spectrum for the six KGB models M1–M6 (colored lines) and for the three reference $w_{0}w_{a}$CDM cosmologies (black lines). Gray points with error bars are the \texttt{CamSpec} bandpower estimates derived from the reanalysis of the Planck PR4 temperature maps~\cite{Efstathiou2021}. The lower panel shows residuals with respect to the DESI~DR2 best-fit $w_{0}w_{a}$CDM prediction. Here we use the conventional rescaling $\mathcal{D}_\ell^{TT} \equiv \ell(\ell+1)C_\ell^{TT}/(2\pi)$. \textit{Right:} CMB lensing convergence power spectrum for the six KGB models M1–M6 (colored lines) and for the three reference $w_{0}w_{a}$CDM cosmologies (black lines). Gray points with uncertainties are the Planck 2018 lensing bandpower measurements obtained using the minimum-variance (MV) estimator over the conservative multipole range $8 \leq \ell \leq 400$ \cite{PlanckCollaboration2019}. The lower panel shows ratios relative to the DESI~DR2 best-fit $w_{0}w_{a}$CDM prediction.}
\end{figure*}

Figure~\ref{fig:cmb} presents the CMB temperature (left) and lensing potential (right) power spectra for the six selected KGB models. For the TT spectrum, we show the \texttt{CamSpec} bandpower estimates derived from the Planck PR4 temperature maps~\cite{Efstathiou2021}, while the lensing panel includes the Planck 2018 lensing bandpower measurements with associated uncertainties~\cite{PlanckCollaboration2019}. All KGB models remain consistent with the observed TT power spectrum across the full multipole range. The lower sub-panel shows residuals relative to the fiducial $w_0w_a$CDM model, revealing that the main differences arise from small shifts in the acoustic peak positions for M5, M6, and the mirage models, driven by their distinct background expansions (see Fig.~\ref{fig:background_perturbations}). Beyond this, the only visible imprint of modified gravity is a mild suppression of the low-$\ell$ ISW plateau, common to all KGB models. The expected increase in peak smoothing due to enhanced lensing potentials is present but negligible—barely perceptible even in the most extreme cases (e.g., M4), and well below the level of experimental noise.

The right panel shows the CMB lensing potential power spectrum, with the lower sub-panel displaying residuals relative to the fiducial $w_0w_a$CDM prediction. All KGB models exhibit a clear increase of lensing power at large angular scales ($\ell \lesssim 100$) compared to their $w_0w_a$CDM counterparts. This results from both the enhanced growth of structure ($\mu > 1$) and a stronger effective gravitational coupling to photons ($\Sigma > 1$), which together boost the amplitude of the lensing potential. Model M5 predicts the largest lensing signal, which may slightly overshoot the data, albeit still within current observational uncertainties.

\begin{figure*}
  \centering
  \includegraphics[width=\textwidth]{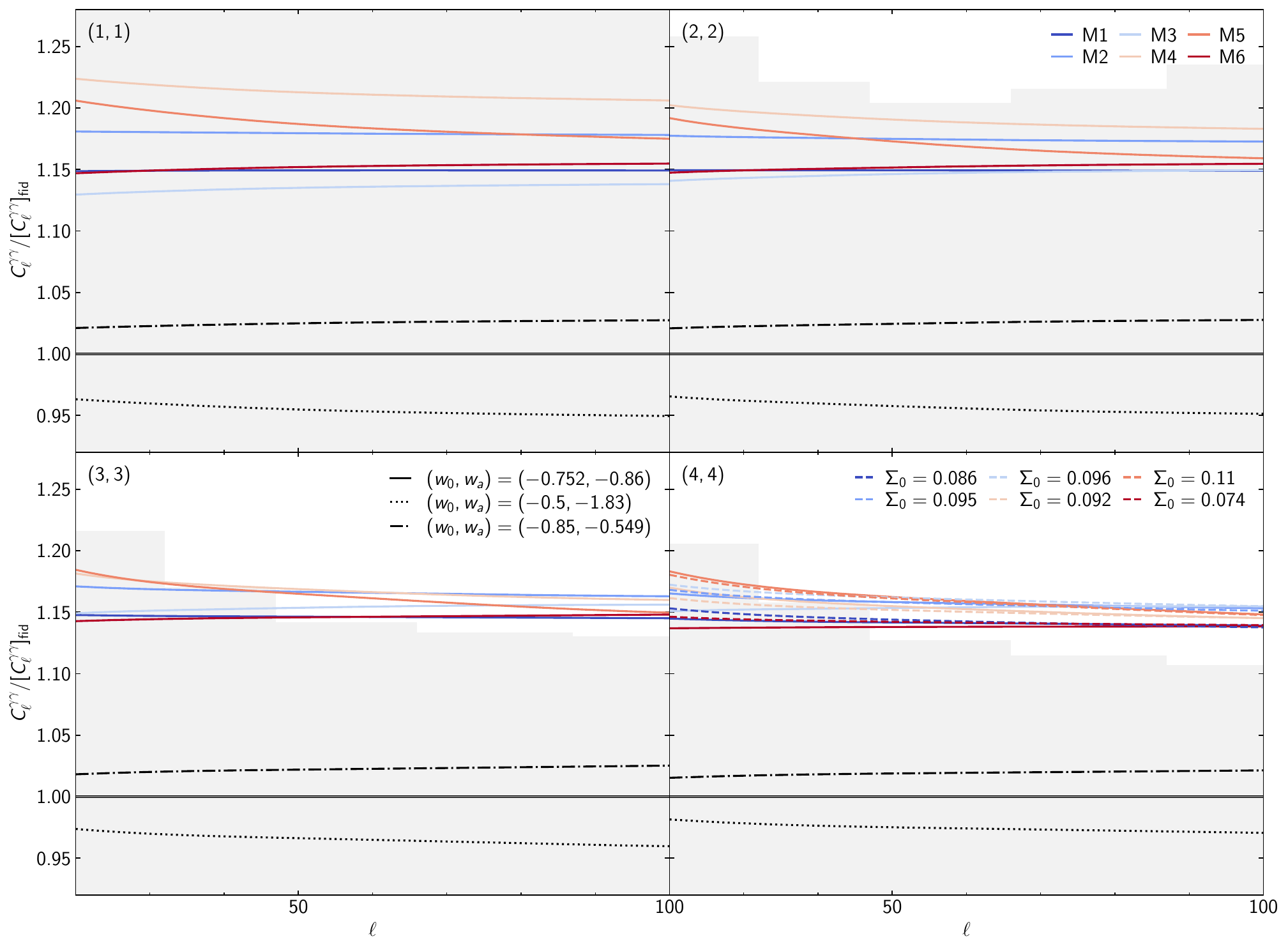}
  \caption{\label{fig:cosmic_shear} Ratios of the cosmic shear auto-power spectra for the four DES Y3 tomographic bins, shown relative to the DESI DR2 best-fit $w_0w_a$CDM prediction. Solid colored curves show the six KGB models M1–M6; black lines correspond to the three reference $w_0w_a$CDM cosmologies; and dashed colored curves in the bottom right panel display the matched phenomenological models parameterized by $\Sigma_0$ (with $\mu_0$ fixed to the values used in Fig.~\ref{fig:fsigma8}). Gray bands indicate the 68\% uncertainties on each tomographic auto-spectrum for a DES Y3-like survey, expressed as $\pm\sigma[C_\ell^{\gamma\gamma}]/C_\ell^{\gamma\gamma}$, including only the Gaussian contributions.}
\end{figure*}

While RSD measurements constrain $\mu$ primarily through its cumulative impact on the growth function and its time derivative, weak lensing is directly sensitive to $\Sigma$ via the lensing kernel in the line-of-sight projection (see Eqs.~\ref{eq:weyl_potential} and \ref{eq:kernel_shear}). Since for the MG cosmologies studied here $\Sigma = \mu$ exceeds unity by 2–12\% across the redshift range probed by galaxy surveys (see Fig.~\ref{fig:background_perturbations}, right panel), the cosmic shear power spectrum emerges as a powerful probe of this class of models, especially when the standard cosmological parameters and expansion history are anchored by CMB, BAO and SNIa data. 

In Fig.~\ref{fig:cosmic_shear}, we show the ratios of the cosmic shear auto-power spectra for the four DES Y3 tomographic bins, computed for the six selected KGB models (solid colored lines) and compared to the DESI DR2 best-fit $w_0w_a$CDM prediction. The gray bands represent the 1$\sigma$ uncertainties on the $i$th tomographic bin, estimated under the Gaussian approximation as (see, e.g., Ref.~\cite{Barreira2018})
\begin{equation}
\sigma^2[C_{\ell,i}^{\gamma\gamma}] = \frac{2}{f_{\rm sky}(2\ell+1)\Delta\ell}\left(C_{\ell,i}^{\gamma\gamma} + \frac{\sigma_e^2}{\bar{n}_{\rm eff}^{i}}\right)^2,
\end{equation}
where $\ell$ denotes the central multipole of each top-hat bin of width $\Delta\ell$. We use $C_{\ell,i}^{\gamma\gamma}$ computed in our fiducial $w_0w_a$CDM cosmology, and assume a DES Y3-like survey~\cite{Abbott2021} with sky fraction $f_{\rm sky} = 0.1$, total RMS ellipticity of the source galaxies $\sigma_e = 0.26$, and effective number density per tomographic bin $\bar{n}_{\rm eff}^{i} = \bar{n}_{\rm eff}/N_{\rm tomo}$, with $\bar{n}_{\rm eff} = 5.9$ arcmin$^{-2}$ and $N_{\rm tomo} = 4$. We neglect super-sample covariance, non-Gaussian terms, and mode-mixing effects from masking and binning. As a result, our uncertainties can be underestimated by up to $\sim 50\%$ compared to the full treatment (cf. Ref.~\cite{Doux2022}).

We adopt the same top-hat binning scheme as in Ref.~\cite{Doux2022} and restrict the comparison to $\ell < 100$, where linear theory provides an accurate description of the $C_\ell^{\gamma\gamma}$ ratios. These scale cuts are motivated by the DES Y3 cosmic shear analysis in modified gravity~\cite{DESCollaboration2023} and are chosen to ensure robustness against nonlinear evolution and baryonic feedback. The effect of modified gravity is clearly visible as an overall enhancement of the lensing signal compared to the $w_0w_a$CDM reference. This trend is consistent with the behavior seen in the CMB lensing spectrum (cf. Fig.~\ref{fig:cmb}). 

In addition to the full KGB predictions, the lower right panel also shows matched phenomenological models (dashed lines), obtained by fixing $\mu_0$ to the values used in Fig.~\ref{fig:fsigma8} and adjusting $\Sigma_0$ to approximate the KGB predictions in the highest tomographic bin, where the signal-to-noise is greatest. The resulting $\mu_0$–$\Sigma_0$ combinations lie within the $95\%$ credibility region derived from the joint DESI DR1 analysis of BAO, RSD, SNIa, CMB, and 3$\times$2pt data in Ref.~\cite{Ishak2025}. A similar level of consistency is found when comparing our models to the more recent KiDS-Legacy analysis by~\cite{Stlzner2025}, particularly the $\{ \Dkin, \cs \}$ model shown in their Fig.~7.

\begin{figure*}
  \centering
  \begin{minipage}[t]{0.5\textwidth}
    \centering
    \includegraphics[height=0.28\textheight]{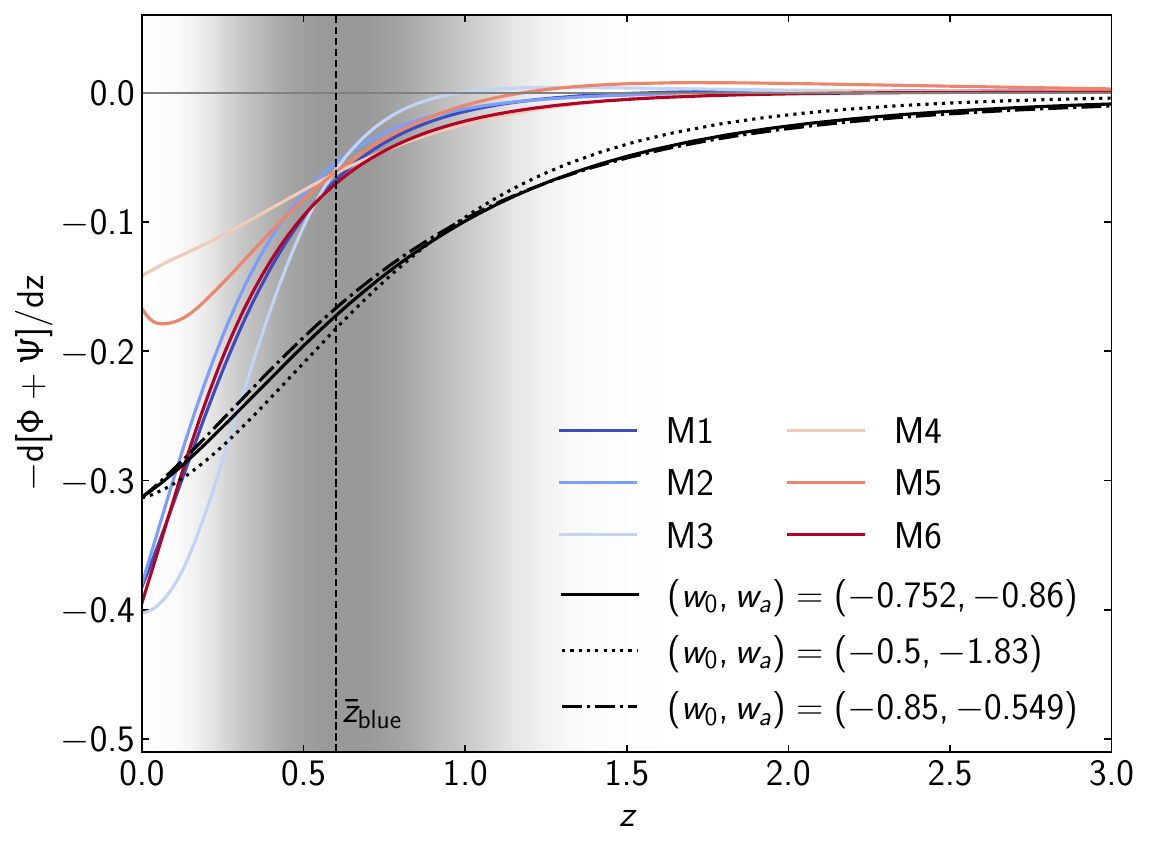}
  \end{minipage}%
  \hfill
  \begin{minipage}[t]{0.5\textwidth}
    \centering
    \includegraphics[height=0.28\textheight]{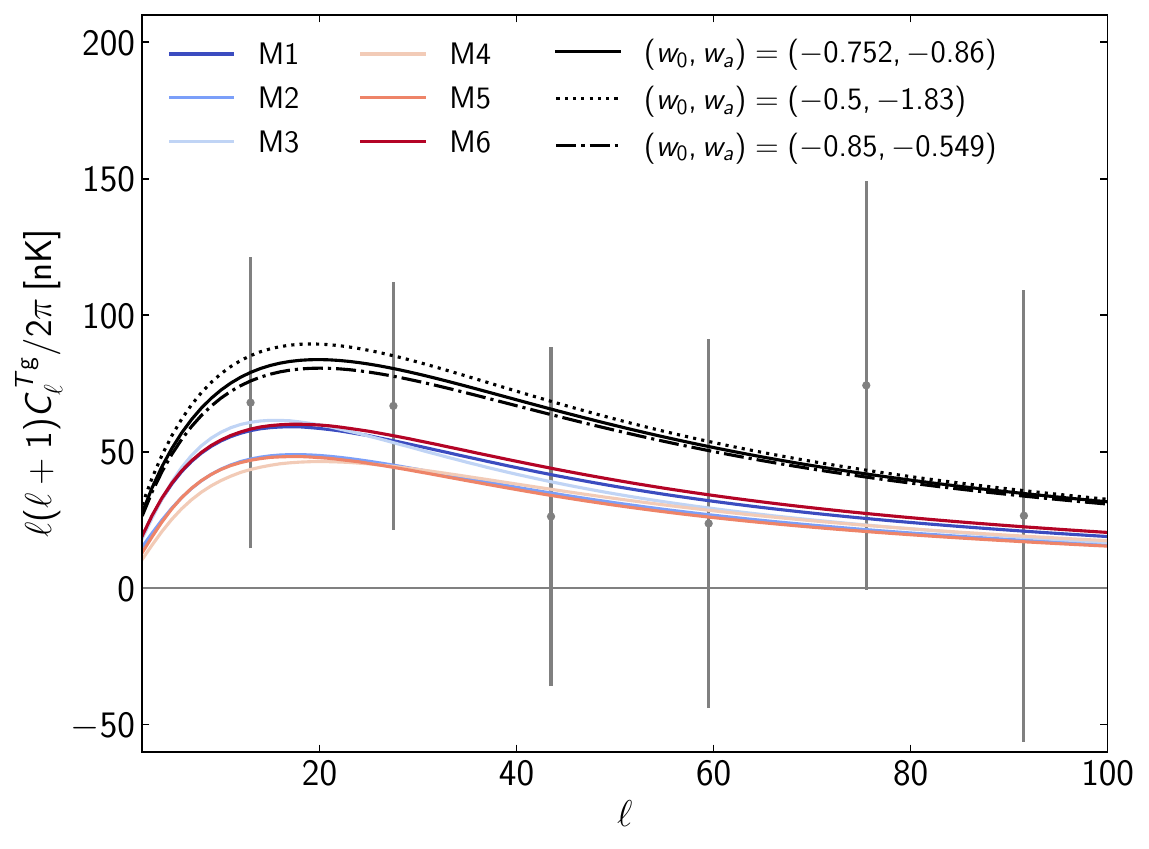}
  \end{minipage}
  \caption{\label{fig:gisw} \textit{Left:} Time derivative of the Weyl potential evaluated at $k \approx 0.01 \, h \, \mathrm{Mpc}^{-1}$, for the six KGB models M1–M6 (colored lines) and the three reference $w_{0}w_{a}$CDM cosmologies (black lines). The grayscale shading encodes the redshift distribution of the unWISE blue galaxy sample~\cite{Krolewski2025}, with the shading intensity proportional to $n_{\rm g}(z)$; the vertical dashed line marks the sample’s mean redshift. \textit{Right:} galaxy-ISW cross–power spectrum for the six KGB models M1–M6 (solid colored lines) and for the three reference $w_{0}w_{a}$CDM cosmologies (black lines). Gray points with error bars show the measured cross-correlation between the unWISE blue galaxy sample and the Planck 2018 CMB temperature maps, as reported in Ref.~\cite{Krolewski2025}.}
\end{figure*}

Finally, we consider the galaxy-ISW cross-power spectrum, a probe of the time evolution of gravitational potentials at late times. The sign and amplitude of the cross-correlation encode the rate of change of the Weyl potential during the onset of cosmic acceleration, making this observable particularly valuable for testing modified gravity. As demonstrated in Ref.~\cite{Renk2017}, models predicting a significantly growing potential are strongly disfavored by the observed positive sign of the cross-correlation signal.

The left panel of Fig.~\ref{fig:gisw} displays the time derivative of the Weyl potential at $k \approx 0.01 \,h\,\mathrm{Mpc}^{-1}$ for the KGB models (colored lines) and the reference $w_0w_a$CDM cosmologies (black lines). Under the Limber approximation, and assuming the fiducial $w_0w_a$CDM background, this scale corresponds to multipole $\ell \approx 20$ when evaluated at the mean redshift of the unWISE blue galaxy sample, $\bar{z}_{\rm blue} = 0.6$ \cite{Krolewski2025}. All models predict a decaying Weyl potential at $z \lesssim 1$, with the KGB cosmologies exhibiting a slower decay than their $w_0w_a$CDM counterparts. This behavior leads to a positive galaxy-ISW cross-correlation, as confirmed by the full calculation shown in the right panel: all six KGB models yield amplitudes in good agreement with observations from Planck 2018 temperature maps cross-correlated with the unWISE blue sample \cite{Krolewski2025}. At higher redshifts (not shown), the signal is expected to vanish for the KGB models but persist in the $w_0w_a$CDM cosmologies, which predict a decaying potential up to $z \sim 3$. However, current measurements using the red and green unWISE samples are limited by lower signal-to-noise and are unlikely to distinguish between these scenarios.

\section{Conclusions and outlook}\label{sec:conclusions}

Recent analyses by the DESI collaboration~\cite{Lodha2025,DESICollaboration2025} have shown a preference for phantom crossing in the dark energy equation of state, which—when interpreted within the framework of Horndeski gravity—has been argued to point toward non-minimal conformal coupling, i.e., $\alpha_M \neq 0$~\cite{Chudaykin2024,Ye2025,Pan2025}. In this work, we have examined whether minimally coupled models—specifically, KGB theories—can provide viable alternatives.

To this end, we have taken advantage of two distinctive features implemented in the Einstein-Boltzmann solver \mochiclass: the use of a manifestly stable EFT basis and the ability to pass arbitrary arrays as input for the EFT functions. These capabilities allow us to use non-parametric EFT functions that guarantee models are free from ghost and gradient instabilities, thereby enabling an efficient exploration of physically viable scenarios. By applying observational selection criteria—limiting departures from $\Lambda$CDM growth and requiring a positive galaxy-ISW cross-correlation—we identified KGB models that remain consistent with a wide range of cosmological data, including background expansion observables, the CMB temperature and lensing spectra, redshift-space distortions, and cosmic shear measurements.

Focusing first on background and CMB observables, we find that the KGB models identified here fit the data as well as non-minimally coupled Horndeski theories. This suggests that the apparent preference for non-minimal coupling in Refs.~\cite{Chudaykin2024,Pan2025} may reflect prior volume effects or limitations inherent in other EFT bases. Although theoretical motivations have been proposed for conformally coupled models as a route to phantom crossing~\cite{Ye2025,Wolf2025b}, our results demonstrate that derivative-coupled models offer an equally compelling explanation for current observations (see also Ref.~\cite{Wolf2025c}). Importantly, these models also avoid observational challenges associated with $\alpha_M \neq 0$, such as stringent bounds on the evolution of the effective Planck mass from local tests of gravity~\cite{Hofmann2018,Burrage2020} and Big Bang Nucleosynthesis~\cite{Alvey2020}. Moreover, KGB models predict vanishing gravitational slip, which implies equality between the lensing and dynamical masses of galaxy clusters. This is consistent with measurements combining X-ray, thermal Sunyaev-Zel'dovich, galaxy dynamics, and lensing observables~\cite{Boumechta2023,Pizzuti2016,Pizzuti2025}, which show no significant deviations from $\gamma = 1$.

Beyond scalar perturbations, the tensor sector of these models also exhibits phenomenologically attractive features. By construction, gravitational waves propagate at the speed of light in KGB models, in agreement with multimessenger constraints from GW170817~\cite{LIGOScientificCollaboration2017}. In addition, the luminosity distance inferred from standard sirens coincides with that from standard candles~\cite{Amendola2018,Lagos2019,Wolf2020}. If future gravitational wave experiments detect no statistically significant discrepancy with measurements based on electromagnetic radiation, they will lend support to theories without non-minimal coupling, thereby making KGB models especially relevant (although see Ref.~\cite{Dalang2019} for a discussion of the interpretation of such measurements). However, KGB models also include cubic Galileon-like terms which may induce instabilities in the presence of gravitational waves for sufficiently large braiding, specifically for $\alpha_{\rm B} \gtrsim 10^{-2}$~\cite{Creminelli2020}. While all the models studied in this work satisfy this condition and would therefore be deemed theoretically unviable under this constraint, possible resolutions involve invoking the breakdown of the EFT or appealing to specific UV completions~\cite{Zumalacarregui2020}. Accepting the constraint at face value implies the need to identify KGB models with $\alpha_{\rm B} \lesssim 10^{-2}$ that still fit current background and large-scale structure data. We defer this exploration to future work, though recent studies employing parametric stable EFT functions suggest that such models may exist~\cite{Stlzner2025}.

A further advantage of the KGB framework is the presence of derivative self-interactions that activate Vainshtein screening~\cite{Deffayet2010}. This mechanism suppresses scalar field gradients in high-density regions, allowing the models to evade stringent gravity constraints on galactic~\cite{Landim2024} and Solar System scales~\cite{Will2014}. Consequently, KGB theories can exhibit cosmologically relevant fifth-force effects while remaining compatible with precision tests of GR in the highly nonlinear regime.

As follow-up study, we will build on the present work by performing full Bayesian analyses of the KGB parameter space to reconstruct the EFT functions and background evolution. This will allow for a more rigorous comparison with data in the linear regime, incorporating the latest CMB, SNIa, BAO, RSD, and cosmic shear measurements. In particular, it will be interesting to compare our findings to those of~\citet{Raveri2019}, who showed that KGB models formulated in the $\alpha$-basis preferred to remain non-phantom in the past and became weakly phantom only at low redshift. This trend runs counter to that identified in Refs.~\cite{Lodha2025,DESICollaboration2025}, suggesting that further insight could be gained by re-examining the KGB class with updated datasets and within an alternative EFT basis.

\begin{acknowledgments}
We are grateful to Erik Rosenberg and Steven Gratton for sharing the \texttt{CamSpec} 12.5HMcl CMB temperature power spectra, and to Alex Krolewski for kindly providing the cross-correlation measurements between unWISE infrared galaxies and the Planck 2018 CMB temperature maps. MC would also like to thank Lucas Porth for helpful discussions on root-finding algorithms. MC acknowledges support from the Max Planck Society and the Alexander von Humboldt Foundation in the framework of the Max Planck-Humboldt Research Award endowed by the Federal Ministry of Education and Research.
KK is supported by STFC grant number ST/W001225/1 and ST/B001175/1.
For the purpose of open access, the authors have applied a Creative Commons Attribution (CC BY) licence to any Author Accepted Manuscript version arising from this work. Supporting research data are available on reasonable request from the authors.

\end{acknowledgments}

\bibliography{references}

\end{document}